\documentclass[pra, twocolumn, floatfix]{revtex4}
\usepackage{graphicx}
\usepackage{amsmath, amsfonts, amssymb, bm}
\usepackage{braket,color}
\begin{document}
\title{Strong-field photoionization in two-center atomic systems}
\author{J. Fedyk}
\altaffiliation{Present address: Physikalisch-Chemisches Institut, Universit\"at Heidelberg, Im Neuenheimer Feld 229, 69120 Heidelberg, Germany}
\author{A. B. Voitkiv}
\author{C. M\"uller}
\affiliation{Institut f\"ur Theoretische Physik I, Heinrich Heine Universit\"at D\"usseldorf, Universit\"atsstr. 1, 40225 D\"usseldorf, Germany}
\date{\today}
\begin{abstract}
Photoionization of an atom $A$ by a strong laser field in the presence of a spatially well-separated neighboring atom $B$ is considered. The laser field frequency is assumed to lie below the ionization potential of atom $A$ and be resonant with a dipole-allowed transition in atom $B$. In this situation, the ionization may occur either directly by multiphoton absorption from the laser field at the first atomic center. Or via an indirect pathway involving two-center electron-electron correlations, where the neighbor atom $B$ is first photoexcited and, afterwards, transfers its energy upon deexcitation radiationlessly to atom $A$. Considering monochromatic as well as bichromatic laser fields, we study various coupling regimes of the photoionization process and identify experimentally accessible parameter domains where the two-center channel is dominant.
\end{abstract}
\maketitle

\section{Introduction}
Starting from the early days of quantum mechanics, photoionization (PI) studies have been paving the way towards an increasingly deep and thorough understanding of the structure and dynamics of matter on a microscopic scale. Nowadays this is accomplished by kinematically complete experiments \cite{COLTRIMS} which allow us to put the most advanced photoionization theories to the test. 

In various PI mechanisms, electron-electron correlations play a crucial role. Well-known examples are single-photon double ionization and resonant PI. The latter process relies on  resonant photoexcitation of an autoionizing state, with subsequent Auger decay. In recent years, a similar kind of ionization mechanism has extensively been studied in systems consisting of two (or more) atoms. Here, a resonantly excited atom transfers its excitation energy radiationlessly via interatomic electron-electron correlations to a neighbouring atom leading to its ionization. This Auger-like decay involving two atomic centers is referred to as interatomic Coulombic decay (ICD) \cite{ICD,ICDres,ICDrev}. It has been observed in a  variety of systems, comprising noble gas dimers \cite{dimers}, clusters \cite{clusters} and water molecules \cite{water}. Similar intersite energy transfer mechanisms occur in slow atomic collisions \cite{Smirnov}, between Rydberg atoms in ultracold quantum gases \cite{Weidemueller} and as F\"orster resonances between chromophores \cite{Forster}.

As a closely related process, we have theoretically studied resonant two-center photoionization (2CPI) in heteroatomic systems, consisting of an atom $A$ and a well-separated atom $B$ of different species \cite{2CPI}. It turns out that this ionization channel can be remarkably strong and can dominate over the usual single-center PI by orders of magnitude. The photon energy was assumed to exceed the ionization potential of atom $A$, rendering the absorption of a single photon already sufficient to promote the electron into the continuum. Such a process was experimentally observed in helium-neon dimers using synchrotron radiation \cite{2CPI_exp}. Calculations on PI in two-atomic systems were also presented in \cite{Kuhn,Perina}. Besides, resonant two-photon ionization in a system of two identical atoms was analyzed \cite{identical}. The influence of a second neighbor atom \cite{3CPI} and collective effects in a multiatom ensemble \cite{NCPI} were studied as well.

With the advent of free-electron lasers, it has become possible to study interatomic autoionization processes also in intense photon fields of high frequency \cite{ICD_FEL}. In particular, time-resolved pump-probe measurements of ICD in neon dimers have been performed, where the autoionizing state was populated by resonant one-photon \cite{time-resolved} or two-photon absorption \cite{two-photon}. Correlated electronic decay process and Penning-type ionization have also been observed in clusters and nanodroplets after irradiation by free-electron lasers \cite{FELclusters,Penning}. Very recently, such interatomic processes were found to occur as well in clusters exposed to non-resonant infrared laser fields of high intensity, where an efficient energy transfer between adjacent electrons may proceed due to Rydberg-state formation in a nanoplasma \cite{clusters2015}. Theoreticians also studied strong-field control of ICD in quantum-dot systems \cite{Bande}.

Motivated by these developments, we generalize in the present paper our consideration of the 2CPI process to electron correlation-driven photoionization in strong laser fields. The laser frequency is assumed to lie below the ionization potential of an atom $A$, which is to be ionized, and to be resonant with a dipole-allowed transition between bound states in a neighboring atom $B$. The absorption of multiple photons from the laser field is thus required to promote the electron into the continuum (see Fig.~1). It is assumed that no bound-state resonances are hit in atom $A$. We shall develop a theoretical description of the two-center ionization process which is based on the strong-field approximation to describe the interaction of the active electron in atom $A$ with the laser field. For the resonant coupling of the laser field to atom $B$, two different cases will be considered, where this coupling is either relatively weak or rather strong. The interatomic interaction will be treated as a perturbation throughout. Monochromatic as well as bichromatic laser fields will be considered, with the focus lying on the latter case. By considering suitable two-center model systems we shall show that the photoionization of atom $A$ can be dominated by the two-center channel for parameters which are experimentally accessible.

Our paper is organized as follows. In Sec.~II a theoretical approach to strong-field PI in two-center systems is developed which is based on the strong-field approximation describing the emitted photoelectron by a Volkov state. We will start with considering monochromatic laser fields (Sec.~II.A) and afterwards treat bichromatic laser fields (Sec.~II.B). In both scenarios, the cases of weak and strong laser-atom coupling will be distinguished and their qualitatively different features revealed. In Sec.~III our theoretical findings are illustrated by some numerical examples, which compare the strengths of various single-center and two-center ionization channels. Our conclusions are given in Sec.~IV. Atomic units (a.u.) are used throughout unless otherwise stated.

\begin{figure}[t]  
\vspace{-0.25cm}
\begin{center}
\includegraphics[width=0.42\textwidth]{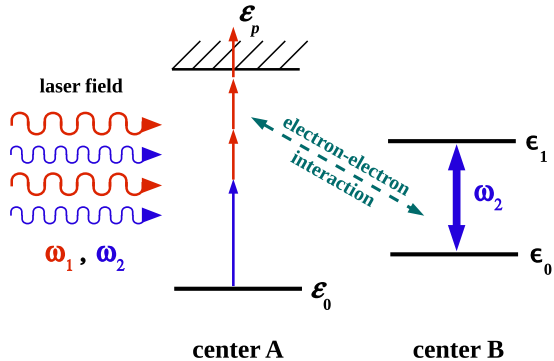}
\end{center}
\vspace{-0.5cm} 
\caption{Scheme of strong-field two-center photoionization in a bichromatic laser field. An assisting atom $B$ is first resonantly photoexcited and, subsequently, transfers the excitation energy to atom $A$. In combination with the energy of several other photons, which are simultaneously absorbed from the laser field, atom $A$ is ionized. If the amplitude of the second frequency mode is rather high, the states of atom $B$ become strongly coupled, as indicated in the picture.}
\label{figure1}
\end{figure}

\section{Theoretical framework}

In order to understand the basic physics of 2CPI in strong laser fields, 
we consider photoionization in a very simple atomic system  
consisting of two atoms ($A$ and $B$) each having just one ``active''
electron. Both are initially in their ground states 
and separated by a distance $R$ large enough, 
that one can still speak about individual atoms.   
Let us further suppose that the atomic nuclei having charge numbers  
$Z_A$ and $Z_B$, respectively, 
are at rest. We shall take the position of the nucleus $Z_A$ 
as the origin and denote the coordinates 
of the nucleus $Z_B$, the electron of the atom $A$ 
and that of the atom $B$ by ${\bf R}$, ${\bf r}$ 
and ${\bf r}' ={\bf R} + \mbox{\boldmath{ $\xi$ }} $, 
respectively, where $\mbox{\boldmath{ $\xi$ }}$ 
is the position of the electron of atom $B$ 
with respect to the nucleus $Z_B$.

\subsection{Monochromatic laser field}
\label{mono}
We first consider two-center photoionization in a monochromatic laser field
${\bm{\mathcal{A}}}(t)$ of frequency $\omega$, which is taken in the dipole approximation. 
For definiteness, the latter is assumed to be linearly polarized, 
\begin{equation}
{\bm{\mathcal{A}}}(t) = A_0 \cos(\omega t) {\bf e}\ .
\label{Alin}
\end{equation}
The corresponding electric field amplitude is $F_0 = \frac{\omega}{c}A_0$. 
Two different limiting cases will be considered. They are distinguished by the relative value of the Rabi frequency $\Omega_B \sim F_0a_0$, which is associated with the dipole transition in atom $B$, as compared with the radiative width $\Gamma_B$ of the excited state in atom $B$. Here, $a_0$ denotes the Bohr radius. First, we shall discuss the case of intense laser fields whose amplitude, still, is small enough that the coupling to atom $B$ may be treated perturbatively ($\Omega_B\ll \Gamma_B$). Afterwards, the opposite case of nonperturbative strong coupling is treated, where the dynamics is determined by the Rabi frequency ($\Omega_B\gg \Gamma_B$).

\subsubsection{Perturbative coupling of the field to atom $B$}
\label{weak}

Two-center ionization involving the absorption of a single photon from such a field
was studied in \cite{2CPI}. In the regime of low field intensities, the process can be 
calculated starting from the second-order amplitude
\begin{eqnarray}
S_{2}^{\rm (m)} &=& i\int_{-\infty}^{\infty} dt \braket{ \Psi_{{\bf p}0}|\hat{V}_{AB}| \Psi_{01} }
e^{-i(E_{01}-E_{p0})t} \nonumber\\
& & \times 
\braket{ \Psi_{01}|\hat{W}_B |\Psi_{00} } \frac{e^{-i(E_{00}-E_{01}+\omega)t}}{E_{00}-E_{01}+\omega+\frac{i}{2}\Gamma_B} \nonumber\\
&=& 2\pi i\, \frac{\braket{ \Psi_{{\bf p}0}|\hat{V}_{AB}| \Psi_{01} }\braket{ \chi_1|\hat{W}_B |\chi_0 }}
{\epsilon_0-\epsilon_1+\omega+\frac{i}{2}\Gamma_B}\,\delta(\varepsilon_0 - \varepsilon_p + \omega) \nonumber\\
& & 
\label{eq:S2}
\end{eqnarray}
Here and henceforth, the superscript ``(m)'' will be used to indicate the monochromatic case. The relevant two-electron configurations involved in this expression are:
(I) $\Psi_{00} = \varphi_0({\bf r}_1) \chi_0(\boldsymbol{\xi})$ with total energy $E_{00} = \varepsilon_0 + \epsilon_0$, describing both electrons in the corresponding ground states $ \varphi_0 $ and $ \chi_0 $; 
(II) $\Psi_{01} = \varphi_0({\bf r}_1) \chi_1(\boldsymbol{\xi})$ with total energy $E_{01} = \varepsilon_0 + \epsilon_1$, in which the electron of the atom $A$ is in the ground state while the electron of the atom $B$ is in the excited state $\chi_1$; and
(III) $\Psi_{{\bf p},0} = \varphi_{\bf p}({\bf r}) \chi_0(\boldsymbol{\xi})$ with total energy $E_{{\bf p}0} = \varepsilon_p + \epsilon_0$, where the electron of the atom $A$ is in a continuum state $\varphi_{\bf p}$ and the electron of the atom $B$ in the ground state. $\Gamma_B$ denotes the radiative width of $\chi_1$.
The photoexcitation of atom $B$ is induced by the interaction term
\begin{eqnarray} 
{\hat W}_B = \frac{A_0}{2c}\,\hat{\bf p}_B\cdot{\bf e}\ ,
\end{eqnarray}
with $\hat{\bf p}_B$ denoting the momentum operator of the electron at center $B$,
whereas the interaction between the atomic transition dipoles is governed by
\begin{eqnarray} 
\hat{V}_{AB} = \frac{ {\bf r}\cdot {\bm \xi} }{ R^3 }
 - \frac{ 3({\bf r}\cdot{\bf R})({\bm \xi}\cdot{\bf R})}{ R^5 }\ .
\label{VAB}
\end{eqnarray}
Accordingly, the first matrix element in Eq.~\eqref{eq:S2} describes the photoexcitation of atom $B$ and the second matrix element describes the interatomic energy transfer, leading to ionization of atom $A$. The $\delta$ function expresses the energy conservation in the process.

The generalization of Eq.~\eqref{eq:S2} to account for the possibility of multiphoton absorption by atom $A$ can be achieved by using the strong-field approximation (SFA) \cite{Friedrich}. The continuum state $\varphi_{\bf p}({\bf r})\,e^{-i\varepsilon_p t}$ is replaced by a Volkov state
\begin{eqnarray}
\psi_{\bf p}^{(\mathcal{A})}({\bf r},t)  = \dfrac{e^{i{\bf p}\cdot{\bf r}}}{\sqrt{V}} \exp \left( -\dfrac{i}{2} \int^{t} [{\bf p}+\dfrac{1}{c}{\bm{\mathcal{A}}}(t')]^2 dt' \right)
\label{eq:vol}
\end{eqnarray}
which is taken in the velocity gauge. Here, $V$ denotes the normalization volume. The corresponding transition amplitude can be written as
\begin{eqnarray}
S_{2}^{\rm (m)} &=& i\int_{-\infty}^{\infty} dt \braket{ \psi_{\bf p}^{(\mathcal{A})}\chi_0|\hat{V}_{AB}| \varphi_0\chi_1 }
e^{-i(\varepsilon_0 + \omega)t} \nonumber\\
& & \times \frac{\braket{ \chi_1|\hat{W}_B |\chi_0 }}{\epsilon_0-\epsilon_1+\omega+\frac{i}{2}\Gamma_B}\ .
\label{eq:S2SFA}
\end{eqnarray}
The time integral can be evaluated by performing a Fourier series expansion based on the generating function of the Bessel functions [see Eq.~\eqref{Bessel} below]. Then, the transition amplitude adopts the form
\begin{eqnarray}
S_{2}^{\rm (m)} &=& \frac{2\pi i}{\sqrt{V}}\sum_{n\ge n_0}^\infty 
\frac{\braket{ e^{i{\bf p}\cdot{\bf r}}\chi_0|\hat{V}_{AB}| \varphi_0\chi_1 }\braket{ \chi_1|\hat{W}_B |\chi_0 }}
{\epsilon_0-\epsilon_1+\omega+\frac{i}{2}\Gamma_B}\nonumber\\
& & \times\, C_n\,\delta(\varepsilon_0 - \varepsilon_p - U_p + n\omega)\ .
\label{eq:S2SFA2}
\end{eqnarray}
Assuming hydrogenlike wavefunctions for the bound states, the spatial intergrations in the matrix elements can be performed by elementary means. The summation index $n$ counts the number of photons absorbed in the process; $n_0$ is the smallest integer with $n\omega + \varepsilon_0 - U_p\ge 0$, such that the argument of the $\delta$ function can be fulfilled. Here, $U_p$ denotes the ponderomotive energy in the laser field. The coefficients $C_n$ generally depend on the field parameters and the electron momentum. For the case of a linearly polarized laser field \eqref{Alin}, the ponderomotive energy reads $U_p = \frac{A_0^2}{4c^2}$ and the coefficients are given by $C_n = \tilde{J}_n(\alpha, \beta)$, where $\tilde{J}_n$ denotes a generalized Bessel function which is related to the ordinary, cylindrical Bessel functions through the identity \cite{Reiss}
\begin{equation}
\tilde{J}_n(\alpha, \beta) = \sum_\ell J_{n-2\ell}(\alpha)J_\ell(\beta)\ ,
\end{equation}
where
\begin{equation}
\alpha = \frac{A_0}{c\omega}\,{\bf p}\cdot{\bf e}\ \ ,\ \ \ \beta = -\frac{A_0^2}{8c^2\omega}\ .
\end{equation}
From the amplitude \eqref{eq:S2SFA2}, the monochromatic two-center ionization rate is obtained by taking the absolute square and integrating over the outgoing electron momenta:
\begin{eqnarray}
\mathcal{R}_{2}^{\rm (m)} = \frac{1}{T}\int \frac{Vd^3p}{(2\pi)^3}\,\left|S_{2}^{\rm (m)}\right|^2\ ,
\label{R2-mono}
\end{eqnarray}
where $T$ denotes the interaction time.

We point out that more advanced SFA theories than the basic one applied in Eq.~\eqref{eq:S2SFA} exist as well (see, e.g., \cite{beyondSFA} and references therein). In this paper, however, our main goal is to reveal the relative importance of strong-field 2CPI as compared with the corresponding well-established single-center ionization process. The ratio of both rates is therefore most relevant for us. Since both rates will be calculated within the same basic SFA formalism, this ratio will be less sensitive to the applied approximation than the separate rates are.

\subsubsection{Nonperturbative coupling of the field to atom $B$}

A laser field which is resonant with the transition in atom $B$ can drive Rabi oscillations between the ground and excited states $\chi_0$ und $\chi_1$. The corresponding Rabi frequency $\Omega_B$ induces a splitting of the level (quasi)energies due to the dynamic Stark effect. If this splitting is larger than the width $\Gamma_B$ due to spontaneous radiative decay, another theoretical description of 2CPI than in Sec.~\ref{weak} is necessary. 

The strong coupling between the laser field and atom $B$ requires a nonperturbative treatment. This can be achieved by using field-dressed bound states instead of the stationary states $\chi_0$ and $\chi_1$ of atom $B$. They can be written as
\begin{eqnarray}
\Phi_{\pm}(\boldsymbol{\xi},t) &=& \left[ (\Delta\mp\Omega_B) e^{i\omega t} \chi_0(\boldsymbol{\xi}) + 2 W_{10}\chi_1(\boldsymbol{\xi}) \right] \nonumber\\
& & \times\,\frac{e^{-i \epsilon_\pm t}}{\sqrt{(\Delta\mp\Omega_B)^2+4|W_{10}|^2}}\ ,
\label{dressed}
\end{eqnarray}
with the detuning $\Delta = \epsilon_0 + \omega - \epsilon_1$, 
the Rabi frequency $\Omega_B = \sqrt{\Delta^2+4|W_{10}|^2}$,
$W_{10} = \braket{\chi_1|W_B|\chi_0}$, and 
$\epsilon_\pm = \frac{1}{2}(\epsilon_0+\epsilon_1+\omega\mp\Omega_B)$.
In the derivation of Eq.~\eqref{dressed}, the rotating wave approximation has been used
and the interaction with the field is assumed to be switched on adiabatically \cite{knight-review, fedorov}. 

The field-dressed states $\Phi_\pm$ are now used as basis states for the combined system ``atom $B$ + laser field''. The only remaining interaction is the interatomic dipole-dipole coupling, which is treated in the first order of perturbation theory, as before. The ionization amplitudes, accordingly, have the form
\begin{equation}
S_{2\pm} = i \int_{-\infty}^{\infty} dt \braket{\psi_{\bf p}^{(\mathcal{A})} \Phi_{\pm}| \hat{V}_{AB} | \varphi_0 \Phi_{+}} e^{-i\varepsilon_0 t}\ .
\label{S+}
\end{equation}
Note that, in our situation, the proper initial condition at $t\to-\infty$ is encoded in the state $\Phi_+$.
Since the interaction of atom $B$ with the laser field has been incorporated in the dressed states, the structure of Eq.~\eqref{S+} looks simpler than the second-order amplitude \eqref{eq:S2SFA}.
Note, however, that there are two partial contributions, $S_{2+}$ and $S_{2-}$, which differ by the final field-dressed state in atom $B$. The spatiotemporal integrations in Eq.~\eqref{S+} can be performed in a straightforward way, resulting in a lengthy expression which is omitted here.

Since the total final states $\psi_{\bf p}^{(\mathcal{A})} \Phi_{+}$ and $\psi_{\bf p}^{(\mathcal{A})} \Phi_{-}$ are orthogonal to each other, the amplitudes in Eq.~\eqref{S+} add up incoherently to yield the ionization rate
\begin{eqnarray}
\mathcal{R}_2^{\rm (m)} = \frac{1}{T}\int \frac{Vd^3p}{(2\pi)^3}\,\left( |S_{2+}|^2+|S_{2-}|^2 \right)\ .
\label{R2+-}
\end{eqnarray}

Before proceeding to the next section, we point out that our consideration of the monochromatic case has been performed mainly for reasons of completeness and for building a bridge to our previous studies of 2CPI by single-photon absorption \cite{2CPI}. Particularly in the case of relatively weak fields  -- where the permissible laser intensities are restricted by the condition $\Omega_B\ll \Gamma_B$ -- the probability for multiphoton absorption is very small (see Sec.~III and also Ref.~\cite{Bande}). As a consequence, our main focus in the present paper shall lie on 2CPI in bichromatic laser fields where these limitations can be circumvented by a proper choice of the field parameters. This more complex case will be considered next.

\subsection{Bichromatic laser field}
Now we turn to two-center photoionization in a bichromatic laser field. The latter is assumed to be of the form
\begin{equation}
{\bm{\mathcal{A}}}(t) = {\bm{\mathcal A}}_1(t) + {\bm{\mathcal A}}_2(t)\ ,
\label{Abi}
\end{equation}
with a strong, low-frequency component of circular polarization
\begin{equation}
{\bm{\mathcal A}}_1(t) = A_{01} \left[ \cos(\omega_1 t) {\bf e}_x + \sin(\omega_1 t) {\bf e}_y \right]
\end{equation}
and a comparably weak, high-frequency component of linear polarization
\begin{equation}
{\bm{\mathcal A}}_2(t) = A_{02} \cos(\omega_2 t) {\bf e}\ ,
\label{A2}
\end{equation}
whose amplitude satisfies the relation $A_{02}\ll A_{01}$. The corresponding field strengths are given by $F_{0j}=\frac{\omega_j}{c}\,A_{0j}$ for $j\in\{1,2\}$. The higher frequency $\omega_2$ is assumed to be resonant with a dipole-allowed transition in atom $B$. We remark that the low-frequency field is chosen to have circular polarization merely for reasons of computational convenience.

In principle, each atom is subject to both fields ${\bm{\mathcal{A}}}_1$ and ${\bm{\mathcal{A}}}_2$. However, we may simplify our treatment considerably by noting that a resonant field can couple two bound states much more efficiently than a nonresonant field, even though the amplitude of the former may be much lower than the amplitude of the latter. Conversely, an intense field of low-frequency exerts a much stronger impact on an electron in the continuum than a weak field of high frequency does. Therefore, to a good approximation, we shall describe the ionized electron by a Volkov state $\psi_{\bf p}^{(\mathcal{A}_1)}({\bf r},t)$ [see Eq.~\eqref{eq:vol}] which includes the strong field ${\bm{\mathcal{A}}}_1$ only. With regard to atom $B$, we will consider only the interaction with the resonant field ${\bm{\mathcal{A}}}_2$. Its (nonresonant) impact on atom $A$ will instead be ignored.

As before, we shall distinguish the cases where the resonant field component is rather weak or relatively strong, respectively, in terms of the relation between the induced Rabi frequency and the radiative line width. We note that, in contrast to the monochromatic laser field of Sec.~\ref{mono}, a bichromatic field offers the advantage that the amplitude of the resonant mode can be kept quite small, without suppressing the probability for multiphoton absorption which can be controlled by the amplitude of the low-frequency mode.

\subsubsection{Weak resonant field}
\label{weak-res}
We first discuss the case, when the high-frequency field component is weak (i.e. $F_{02}a_0\sim\Omega_B\ll \Gamma_B$). Then, its interaction with atom $B$ can be treated in the first order of perturbation theory. Accordingly, the ionization amplitude in the bichromatic field can be written approximately as
\begin{eqnarray}
S_{2}^{\rm (bi)} &=& i\int_{-\infty}^{\infty} dt \braket{ \psi_{\bf p}^{(A_1)}\chi_0|\hat{V}_{AB}| \varphi_0\chi_1 } e^{-i(\varepsilon_0 + \omega_2)t} \nonumber\\
& & \times \frac{\braket{ \chi_1|\hat{W}_B^{(2)} |\chi_0 }}{\Delta+\frac{i}{2}\Gamma_B}\ ,
\label{eq:S2SFAbi}
\end{eqnarray}
where $\hat{W}_B^{(2)} = \frac{A_{02}}{2c}\,\hat{\bf p}_B\cdot{\bf e}$ and $\Delta = \epsilon_0-\epsilon_1+\omega_2$ is the detuning. Besides, the superscript ``(bi)'' is used to indicate the bichromatic case. Similarly as before, the time integral can be evaluated by performing a Fourier series expansion of the periodic parts in the Volkov states, based on the the identity 
\begin{equation}
e^{i\alpha \sin\eta} = \sum_{n=-\infty}^\infty J_n(\alpha)\,e^{in\eta}\ .
\label{Bessel}
\end{equation}
Then, the bichromatic ionization amplitude becomes
\begin{eqnarray}
S_{2}^{\rm (bi)} &=& \frac{2\pi i}{\sqrt{V}}\sum_{n\ge n_0}^\infty 
\frac{\braket{ e^{i{\bf p}\cdot{\bf r}}\chi_0|\hat{V}_{AB}| \varphi_0\chi_1 }\braket{ \chi_1|\hat{W}_B^{(2)} |\chi_0 }}
{\Delta+\frac{i}{2}\Gamma_B}\nonumber\\
& & \times\, D_n\,\delta(\varepsilon_0 - \varepsilon_p - U_p + n\omega_1 + \omega_2)\ .
\label{eq:S2SFAbi2}
\end{eqnarray}
Since the strong field is circularly polarized, the coefficients are given by ordinary Bessel functions, according to $D_n = J_n(\alpha)\,e^{in\eta_p}$. Here, the argument reads $\alpha = \frac{A_{01}p_\perp}{c\omega_1}$, with $p_\perp = \sqrt{p_x^2+p_y^2}$ denoting the magnitude of the electron momentum component which lies in the polarization plane of the field ${\bm{\mathcal{A}}}_1$, and the phase $\eta_p$ is defined by the relations $p_x = p_\perp\cos\eta_p$ and $p_y = p_\perp\sin\eta_p$. 
The summation index $n$ counts the number of low-frequency photons absorbed from the field ${\bm{\mathcal{A}}}_1$; $n_0$ is the smallest integer with $n\omega_1 + \omega_2 + \varepsilon_0 - U_p\ge 0$. The ponderomotive energy results from the strong, circular-polarized field and reads $U_p = \frac{A_{01}^2}{2c^2}$.

The $\delta$ function in Eq.~\eqref{eq:S2SFAbi2} reflects the law of energy conservation in the process. It shows that the ionization is achieved by combining the energy of a high-frequency photon $\omega_2$, which has first been absorbed by atom $B$, and a variable number $n$ of low-frequency photons $\omega_1$.

From the amplitude \eqref{eq:S2SFAbi2}, we obtain the corresponding ionization rate $\mathcal{R}_{2}^{\rm (bi)}$ by an analogous expression like in Eq.~\eqref{R2-mono}. It can be cast into the following form
\begin{equation}
\mathcal{R}_{2}^{\rm (bi)} = \frac{A_{02}^2\,\Gamma_B^2}{R^6\left[\Delta^2+\frac{1}{4}\Gamma_B^2\right]} \sum_{n\ge n_0}^\infty \mathcal{F}_n
\label{R2_bi_weak}
\end{equation}
which highlights its overall structure and main dependencies. Further more detailed information such as the interatomic geometry are encoded in the functions $\mathcal{F}_n$.

\subsubsection{Strong resonant field}
Let us now turn to the case of a relatively strong high-frequency field component ${\bm{\mathcal{A}}}_2$, satisfying $\Omega_B \gg \Gamma_B$ (but still $A_{02}\ll A_{01}$). In the same spirit as in the previous Sec. \ref{weak-res}, we may obtain the corresponding contributions to the ionization amplitude by performing in Eq.~\eqref{S+} the replacement $\psi_{\bf p}^{(\mathcal{A})}({\bf r},t)\to\psi_{\bf p}^{(\mathcal{A}_1)}({\bf r},t)$ and taking states $\Phi_{\pm}(\boldsymbol{\xi},t)$ in atom $B$ which are dressed by the resonant field ${\bm{\mathcal{A}}}_2$ only. Thus, the action of the field ${\bm{\mathcal{A}}}_2$ is neglected on atom $A$, whereas the action of the field ${\bm{\mathcal{A}}}_1$ is neglected on atom $B$.

Accordingly, the two-center ionization in the bichromatic field is described by the transition amplitudes
\begin{equation}
S_{2\pm} = i \int_{-\infty}^{\infty} dt \braket{\psi_{\bf p}^{(\mathcal{A}_1)} \Phi_{\pm}| \hat{V}_{AB} | \varphi_0 \Phi_{+}} e^{-i\varepsilon_0 t}\ .
\label{S_2C_bi_strong}
\end{equation}
They give rise to two incoherent contributions to the total 2CPI rate, $\mathcal{R}_{2}^{\rm (bi)}=\mathcal{R}_{2+}^{\rm (bi)} + \mathcal{R}_{2-}^{\rm (bi)}$, in analogy with Eq.~\eqref{R2+-}. Focusing on their main dependencies, these partial rates can be written in the form
\begin{equation}
\mathcal{R}_{2+}^{\rm (bi)} = \frac{A_{02}^2\,(\Delta-\Omega_B)^2}{R^6\left[(\Delta-\Omega_B)^2+4|W_{10}|^2\right]^2} \sum_{n\ge n_0}^\infty \mathcal{G}_n
\label{R2_bi_strong}
\end{equation}
and a similar expression holding for $\mathcal{R}_{2-}^{\rm (bi)}$. Here, the functions $\mathcal{G}_n$ contain all remaining dependencies. 

From the structure of Eq.~\eqref{R2_bi_strong} we see that, exactly on the resonance, the rate becomes independent of $A_{02}$. Thus, as a function of the resonant field amplitude, the bichromatic 2CPI rate first increases like $A_{02}^2$ in the weak-coupling regime where $\Omega_B\ll\Gamma_B$ [see Eq.~\eqref{R2_bi_weak}], then the growth reduces in the intermediate transition regime ($\Omega_B\approx\Gamma_B$), and eventually saturation occurs in the strong-coupling regime ($\Omega_B\gg\Gamma_B$).

\subsection{Competing single-center processes}
Below, we shall illustrate the relevance of two-center ionization in a bichromatic laser field by way of several examples. Before doing so, however, we should note that atom $A$ can also be ionized directly, i.e. without participation of atom $B$. There are various channels for this single-center ionization which compete with the 2CPI. If they are too strong, they can mask the two-center ionization. 

In accordance with the usual strong-field approximation in the velocity gauge, the direct ionization of atom $A$ in the bichromatic laser field \eqref{Abi} can be described by the amplitude
\begin{equation}
S_1 = -i \int_{-\infty}^{\infty} dt \braket{ \psi_{\bf p}^{(\mathcal{A}_1+\mathcal{A}_2)} | \hat{H}_{\rm int} | \varphi_0}e^{-i\varepsilon_0 t} \ ,
\label{S1}
\end{equation}
where $\psi_{\bf p}^{(\mathcal{A}_1+\mathcal{A}_2)}$ denotes a Volkov state in the bichromatic field and 
the interaction Hamiltonian
\begin{equation}
\hat{H}_{\rm int} = \hat{H}^{(0)} + \hat{H}^{(1)} + \hat{H}^{(2)}
\label{H_int}
\end{equation}
may be decomposed into contributions of increasing order in the weak-field component:
\begin{equation}
\hat{H}^{(0)} = \frac{1}{c}{\bm{\mathcal{A}}}_1\cdot\hat{\bf p}_A + \frac{1}{2c^2}{\bm{\mathcal{A}}}_1^2
\end{equation}
\begin{equation}
\hat{H}^{(1)} = \frac{1}{c}\left( \hat{\bf p}_A + \frac{1}{c}{\bm{\mathcal{A}}}_1 \right)\cdot{\bm{\mathcal{A}}}_2
\end{equation}
\begin{equation}
\hat{H}^{(2)} = \frac{1}{2c^2}{\bm{\mathcal{A}}}_2^2
\end{equation}
The bichromatic Volkov state can be written as
\begin{eqnarray*}
\psi_{\bf p}^{(\mathcal{A}_1+\mathcal{A}_2)} &=& \psi_{\bf p}^{(\mathcal{A}_1)}\,
e^{-i\int \left[ \frac{1}{c}\left( {\bf p} + \frac{1}{c}{\bm{\mathcal{A}}}_1 \right)\cdot{\bm{\mathcal{A}}}_2 + \frac{1}{2c^2}{\bm{\mathcal{A}}}_2^2\right]dt}\\
&\approx& \psi_{\bf p}^{(\mathcal{A}_1)}\left[ 1 - i\int \frac{1}{c}\left( {\bf p} + \frac{1}{c}{\bm{\mathcal{A}}}_1 \right)\cdot{\bm{\mathcal{A}}}_2\,dt \right]
\end{eqnarray*}
where in the final step an expansion in powers of $\mathcal{A}_2$ has been performed and terms of order $\mathcal{O}(\mathcal{A}_2^2)$ and higher have been dropped.

These decompositions allow us to identify various single-center ionization mechanisms in the amplitude \eqref{S1}. First, it contains the amplitude for ionization of atom $A$ by the strong field ${\bm{\mathcal{A}}}_1$ alone,
\begin{equation}
S_1^{(0)} = -i \int_{-\infty}^{\infty} dt \braket{ \psi_{\bf p}^{(\mathcal{A}_1)} | \hat{H}^{(0)} | \varphi_0}e^{-i\varepsilon_0 t} \ .
\label{S1-0}
\end{equation}
Besides, there is a combined amplitude for ionization which involves the strong field to all orders along with one photon from the weak field, 
\begin{eqnarray}
S_1^{(1)} &=& -i \int_{-\infty}^{\infty} dt \braket{ \psi_{\bf p}^{(\mathcal{A}_1)} | \hat{H}^{(1)}_{\rm eff} | \varphi_0}e^{-i\varepsilon_0 t}\ ,
\label{S1-1}
\end{eqnarray}
with 
\begin{equation}
\hat{H}^{(1)}_{\rm eff} = \hat{H}^{(1)} + \frac{i}{c}\int \left( {\bf p} + \frac{1}{c}{\bm{\mathcal{A}}}_1 \right)\cdot{\bm{\mathcal{A}}}_2\,dt\, \hat{H}^{(0)}\ .
\end{equation}
In analogy with Eq.~\eqref{R2-mono}, the corresponding single-center ionization rates are obtained from 
\begin{eqnarray}
\mathcal{R}_{1}^{(\ell)} = \frac{1}{T}\int \frac{Vd^3p}{(2\pi)^3}\,\left|S_{1}^{(\ell)}\right|^2\ ,
\label{R1}
\end{eqnarray}
with the upper index $\ell\in\{0,1\}$ denoting the order of $\mathcal{A}_2$ being involved.
We emphasize that no quantum interferences between the amplitudes $S_1^{(\ell)}$ arise, provided the frequencies $\omega_1$ and $\omega_2$ are incommensurate.

Ionization pathways which involve the field ${\bm{\mathcal{A}}}_1$ together with higher orders of the field ${\bm{\mathcal{A}}}_2$ will not be considered in the subsequent discussion. For the chosen parameters, they can be estimated to give just a small contribution to the single-center ionization. 
Nevertheless, in our comparative discussion below, we shall include ionization {\it solely} by the field ${\bm{\mathcal{A}}}_2$. It may be calculated approximately from the SFA amplitude
\begin{equation}
S_1^{(2)} = -i \int_{-\infty}^{\infty} dt \braket{ \psi_{\bf p}^{(\mathcal{A}_2)} | 
\frac{1}{c}\hat{\bf p}_A \cdot{\bm{\mathcal{A}}}_2 + \hat{H}^{(2)} | \varphi_0}e^{-i\varepsilon_0 t} \ .
\label{S1-2}
\end{equation}
The corresponding rate $\mathcal{R}_{1}^{(2)}$, which follows from an expression analogous to Eq.~\eqref{R1}, will serve us as a reference value for comparisons with the other ionization mechanisms.

Before moving on to the results section, two comments are appropriate. First, it is possible to embed the two-center ionization amplitude \eqref{S_2C_bi_strong} and the single-center ionization amplitude \eqref{S1} into a common frame, which makes their connection more transparent. Ionization of atom $A$ in a two-center system, which is subject to the bichromatic field \eqref{Abi}, can occur either through its coupling to the field via the Hamiltonian \eqref{H_int} or through the interatomic mechanism involving the dipole interaction \eqref{VAB}. The combined amplitude may thus be written as
\begin{equation*}
S_{12} = i \int_{-\infty}^{\infty} dt \braket{\psi_{\bf p}^{(\mathcal{A}_1+\mathcal{A}_2)} \Phi_f| \big( \hat{H}_{\rm int} + \hat{V}_{AB} \big) | \varphi_0 \Phi_{+}} e^{-i\varepsilon_0 t}\ ,
\label{S_joint}
\end{equation*}
where either $f=+$ or $f=-$. Since $\hat{H}_{\rm int}$ acts on atom $A$ only, this amplitude can be decomposed according to 
\begin{eqnarray*}
S_{12} &=& i\, \delta_{f+} \int_{-\infty}^{\infty} dt \braket{\psi_{\bf p}^{(\mathcal{A}_1+\mathcal{A}_2)} | \hat{H}_{\rm int}| \varphi_0} e^{-i\varepsilon_0 t}\\ 
& & +\,i \int_{-\infty}^{\infty} dt \braket{\psi_{\bf p}^{(\mathcal{A}_1+\mathcal{A}_2)} \Phi_{\pm}| \hat{V}_{AB} | \varphi_0 \Phi_{+}} e^{-i\varepsilon_0 t}\ ,
\end{eqnarray*}
where $\delta_{f+} = \langle\Phi_f|\Phi_{+}\rangle$. Hence, due to the orthogonality of the field-dressed states, the first line of this equation contributes only for $f=+$, and then it coincides with the single-center amplitude \eqref{S1}. The second line gives the two-center amplitude \eqref{S_2C_bi_strong}, where the additional approximation $\psi_{\bf p}^{(\mathcal{A}_1+\mathcal{A}_2)}\approx \psi_{\bf p}^{(\mathcal{A}_1)}$ has been applied because, here, the coupling of the resonant high-frequency field ${\bm{\mathcal{A}}}_2$ to the bound states of atom $B$ is much more relevant than its impact on the continuum state of atom $A$. While this consideration shows that the single-center processes of the current section can be treated with 2CPI in a unified way, their separate calculation offers the advantage that the relative importance of the various ionization mechanisms can be compared with each other (see Sec.~III).

Second, it is worth mentioning that also atom $B$ can be ionized in the presence of the bichromatic field, for example, through resonant ionization by two-photon absorption from the high-frequency mode. This kind of single-center process, however, is well known in the literature (see, e.g., \cite{knight-review}) and not within the scope of the present paper. We are solely interested in the ionization of atom $A$. Therefore, it only matters to us that the ionization of the atoms $B$ is not too strong, so that their majority survives and can participate in the 2CPI of atom $A$. Note that, in an experiment, electrons ejected from atom $B$ can be distinguished from those originating from atom $A$ by their different kinetic energies.

\section{Results and Discussion}

We shall illustrate the results obtained in the previous section by some examples. Our general intention is to see whether two-center ionization in bichromatic laser fields can be a relevant ionization pathway in comparison with the competing processes. To this end we shall consider simplified, generic model systems for the two-center atomic system. Each center is treated as an effective one-electron atom, which is parametrized by an effective nuclear charge $Z_A$ and $Z_B$, respectively. The charges will be chosen in such a way to offer some similarity with real atomic species. The interatomic displacement vector is always taken along the $z$ axis, ${\bf R}=R{\bf e}_z$.

In our first model system we assume that a hydrogen atom represents center $B$. During 2CPI the $1s\to2p$ transition with $\epsilon_1-\epsilon_0 \approx 10.2$\,eV is resonantly driven. The partner atom $A$ is supposed to have an ionization potential which is larger than the excitation energy, but smaller than the binding energy in hydrogen. These conditions guarantee that (i) atom $A$ cannot be ionized by single-photon absorption from the resonant field and that (ii) it is somewhat easier to ionize atom $A$ than atom $B$, since the latter process would reduce the number of two-center systems which can contribute to 2CPI. We chose an ionization potential of $|\varepsilon_0|\approx 12.1$\,eV, corresponding to $Z_A=0.94$. For simplicity, the ground state of atom $A$ is assumed to be describable by a $1s$ wavefunction. A very simple prototype model for a two-center system is established this way. To have a succinct name, we will denote the system as ``Xe-H-like'' since the chosen ionization potential coincides with the value in xenon. 

The parameters of the second model are chosen to mimick a really existing system more closely, taking a He-Ne dimer as a reference. Helium represents the atom $A$, which is to be ionized from the ground state; the effective nuclear charge is chosen as $Z_A=1.345$ to match the binding energy $|\varepsilon_0|\approx 24.6$\,eV of helium. To model neon as the neighboring atom $B$, the resonant photoexcitation is calculated from a $2p$ to a $3s$ state, with $Z_B=1.259$ chosen in correspondence with the excitation energy $\epsilon_1-\epsilon_0 \approx 16.85$\,eV in neon \cite{NIST}. Our ``He-Ne-like'' model system thus captures some basic features of a real He-Ne dimer. Note that this van-der-Waals molecule was used in the experimental studies of 2CPI \cite{2CPI_exp}. In its electronic ground state, the interatomic distance varies between $R\approx 2$--8\,\AA, with the minimum of the potential curve lying at the equilibrium distance $R_{\rm eq}\approx 3$\,\AA\ \cite{Sisourat}.

The frequency $\omega_2$ of the high-frequency field mode is always chosen to be in exact resonance with the transition energy in atom $B$, which is lower than the binding energy in atom $A$. Thus, in contrast to our previous studies in \cite{2CPI} and the experiment in \cite{2CPI_exp}, the absorption of more than one photon is required. The ionization potential of atom $A$ can be surmounted either by absorption of two (or more) high-frequency photons $\omega_2$. Or in a genuinly bichromatic process by absorbing one high-frequency photon together with a number of low-frequency photons. The parameters in our model systems are chosen such that a single high-frequency photon provides already a large fraction of the required energy. The low frequency is supposed to satisfy the condition $\omega_1\ll\omega_2$. Besides, it is assumed that no resonance is hit in atom $A$ while climbing the energy ladder $-\varepsilon_0+n\omega_1+\omega_2$ ($n=0,1,2,\ldots$) up to the continuum. For definiteness, the polarization vector $\bf e$ of the exciting laser field \eqref{A2} will be taken along the $x$ axis, throughout.

We mention that single-center photoionization in a bichromatic laser field with $\omega_2 \lesssim |\varepsilon_0|$ was studied theoretically in \cite{Kalman}. Photoionization of single atoms and ions in bichromatic laser fields with $\omega_1, \omega_2 < |\varepsilon_0|$ was calculated more recently in \cite{Ishikawa}. Several experiments on strong-field photoionization of atoms in bichromatic laser fields were conducted, combining extreme ultraviolet (xuv) or soft x-ray radiation with infrared or optical laser beams. For example, nonresonant photoionization of argon atoms by xuv high-harmonics in the frequency range $17\,\mbox{eV}\lesssim\omega_2\lesssim 38\,\mbox{eV}$ and an intense, near-optical laser pulse ($\omega_1\approx 1.5$\,eV, $I\sim 10^{12}$\,W/cm$^2$) was observed \cite{Meyer2004a}. Related studies applied high-frequency radiation from a synchrotron source \cite{Meyer2004b} or free-electron laser \cite{Meyer2006} in combination with optical laser fields.

\subsection{Weak resonant field}

\begin{figure}[t]  
\vspace{-0.25cm}
\begin{center}
\includegraphics[width=0.5\textwidth]{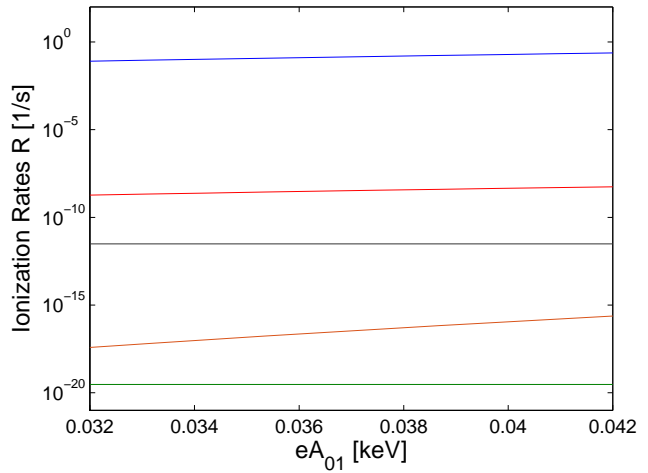}
\end{center}
\vspace{-0.5cm} 
\caption{Photoionization in a Xe-H-like system 
with $\omega_1=1.8$\,eV, $\omega_2=10.2$\,eV, and $eA_{02}=3.7\times 10^{-6}$\,eV (weak-field case) 
at interatomic distance $R=10\,\mbox{\AA}$. Various ionization channels are shown (from top to bottom): bichromatic 2CPI [blue line, see Eq.~\eqref{eq:S2SFAbi2}], bichromatic single-center PI [red line, see Eq.~\eqref{S1-1}], monochromatic 2CPI in the field $\mathcal{A}_2$ [black line, see Eq.~\eqref{R2-mono}], monochromatic single-center PI in the field $\mathcal{A}_1$ [orange line, see Eq.~\eqref{S1-0}], and monochromatic single-center PI in the field $\mathcal{A}_2$ [green line, see Eq.~\eqref{S1-2}].}
\label{figure2}
\end{figure}

Figure~\ref{figure2} shows the rates of various ionization channels in our Xe-H-like two-center model system. The high-frequency field component has a field strength of $F_{02} = 1.9$\,V/cm, corresponding to a Rabi frequency of $\Omega_B\sim3\times 10^{-10}$ a.u. which is by two orders of magnitude smaller than the radiative decay width $\Gamma_B=1.5\times10^{-8}$ a.u. of the excited $2p$ state in hydrogen. The amplitude of the low-frequency field component is varied in Fig.~\ref{figure2}; it has a field strength of the order $F_{01}\sim 3.5\times 10^6$\,V/cm, corresponding to an intensity of $\sim 10^{10}$\,W/cm$^2$.

Despite its small amplitude, the presence of the high-frequency field leads to a strong enhancement of ionization, both for the single-center processes (see red and orange curves) and the two-center processes (see blue and black curves). The high-frequency field alone is not powerful, though, as it leads to a negligibly small ionization rate (green curve). 
For the chosen parameters, the bichromatic 2CPI rate $\mathcal{R}_2^{({\rm bi})}$ [cf. Eq.~\eqref{eq:S2SFAbi2}] is by far the largest. Since the ionization requires two low-frequency photons $\omega_1$ to be absorbed along with one high-frequency photon $\omega_2$, it scales with $\sim A_{01}^4$ and exceeds both the bichromatic single-center ionization rate $\mathcal{R}_1^{(1)}$ [cf. Eq.~\eqref{S1-1}] as well as the monochromatic 2CPI rate $\mathcal{R}_2^{({\rm m})}$ [cf. Eq.~\eqref{R2-mono}] in the resonant field ${\bm{\mathcal{A}}}_2$ by several orders of magnitude. The latter rate relies on the absorption of two high-frequency photons $\omega_2$, in total.

The monochromatic 2CPI rate lies eight orders of magnitude above the monochromatic single-center rate $\mathcal{R}_1^{(2)}$ in the field ${\bm{\mathcal{A}}}_2$ [cf. Eq.~\eqref{S1-2}]. In our previous studies \cite{2CPI}, a ratio between the two-center and corresponding single-center PI rates of roughly $[c/(\omega_2 R)]^6$ was found, which is in good agreement with the current data.

To ionize atom $A$ solely by absorption from the low-frequency field mode, at least $n_0=7$ photons $\omega_1$ are required to overcome the ionization potential. The corresponding monochromatic ionization rate $\mathcal{R}_1^{(0)}$ [cf. Eq.~\eqref{S1-0}] in Fig.~\ref{figure2} scales approximately with $\sim A_{01}^{15}$, indicating that the main contributions stem from $n=7$ and 8 photons.
Note in this context that the ponderomotive energy $U_p\sim 1$\,meV is very small. Accordingly, the Keldysh parameter $\gamma=\sqrt{|\varepsilon_0|/2U_p}\sim 80$ is large and the ionization occurs in the perturbative multiphoton regime.

\begin{figure}[t]  
\vspace{-0.25cm}
\begin{center}
\includegraphics[width=0.5\textwidth]{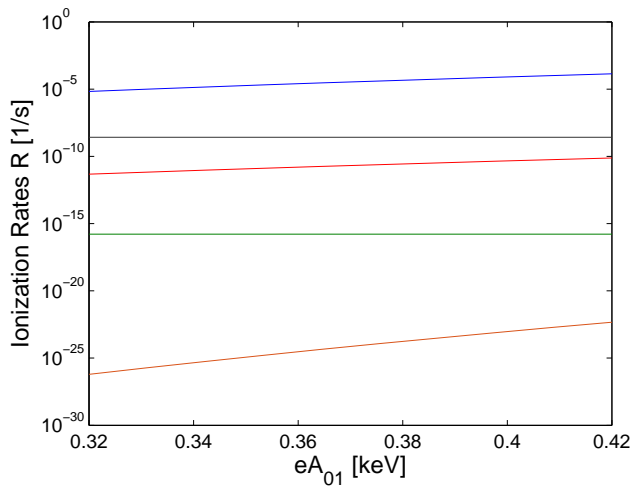}
\end{center}
\vspace{-0.5cm} 
\caption{Photoionization in a He-Ne-like system 
with $\omega_1=1.7$\,eV, $\omega_2=16.85$\,eV, and $eA_{02}=3.7\times 10^{-5}$\,eV (weak-field case) 
at interatomic distance $R=5\,\mbox{\AA}$. The various ionization channels are distinguished by the 
same color coding as in Fig.~\ref{figure2}.}
\label{figure3}
\end{figure}

Our second model system is shown in Fig.~\ref{figure3}. Since the binding energy of helium is twice as large as in xenon, the applied vector potentials are chosen to be larger by one order of magnitude than before, corresponding to field strengths $F_{01}\sim 3.2\times 10^7$\,V/cm and $F_{02}\sim 30$\,V/cm, respectively. As before, the Keldysh parameter $\gamma\sim 35$ indicates perturbative multiphoton ionization and $\Omega_B\ll\Gamma_B$ implies weak coupling between atom $B$ and the resonant field component. The interatomic distance is chosen to lie in the middle of the relevant range mentioned above.

The bichromatic 2CPI rate $\mathcal{R}_2^{({\rm bi})}$ is again the largest. Ionization via this channel requires at least five low-frequency photons $\omega_1$ in addition to one high-frequency photon $\omega_2$. To a good approximation, the rate shows a scaling with $\sim A_{01}^{11}$. It exceeds the bichromatic single-center PI rate $\mathcal{R}_1^{(1)}$ by more than six orders of magnitude. In comparison with Fig.~\ref{figure2} we see that the monochromatic 2CPI channel has become relatively more important. This is because, even though both field amplitudes are enlarged, the probability to absorb a second high-frequency photon has grown more strongly than the probability to absorb the required photons from the low-frequency field, since their number has increased. For similar reasons, the monochromatic PI in the high-frequency field $\mathcal{A}_2$ is much stronger than the monochromatic PI in the low-frequency field $\mathcal{A}_1$. The latter requires absorption of a large number of 15 photons and is, thus, heavily suppressed in the multiphoton regime.

\subsection{Strong resonant field}

\begin{figure}[b]  
\vspace{-0.25cm}
\begin{center}
\includegraphics[width=0.5\textwidth]{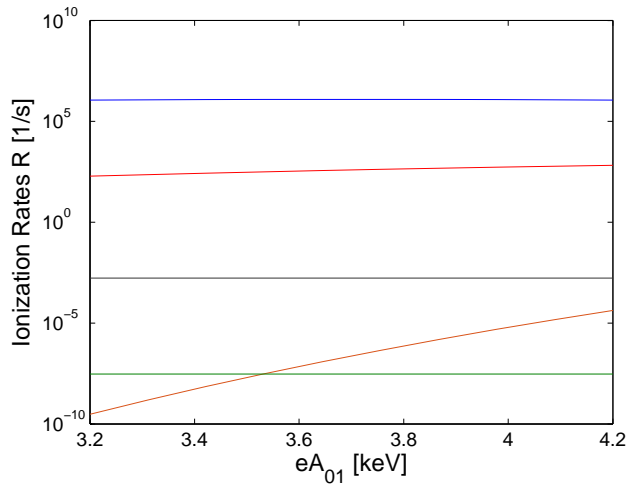}
\end{center}
\vspace{-0.5cm} 
\caption{Photoionization in a Xe-H-like system with $\omega_1=0.3$\,eV, $\omega_2=10.2$\,eV, and $eA_{02}=3.7\times 10^{-3}$\,eV (strong-field case) at interatomic distance $R=10\,\mbox{\AA}$. The various ionization channels are distinguished by the same color coding as in Fig.~\ref{figure2}.}
\label{figure4}
\end{figure}

We now turn to the case of strong laser-atom coupling with $\Omega_B\gg\Gamma_B$. Figure~\ref{figure4} shows our corresponding results for the first model system. For the chosen amplitude of the resonant field component, the Rabi frequency is larger by an order of magnitude than the radiative width. 
As compared with Fig.~\ref{figure2}, the 2CPI rates $\mathcal{R}_2^{({\rm bi})}$ and $\mathcal{R}_2^{({\rm m})}$ have increased by several orders of magnitude, accordingly. Note, however, that the increase of $\mathcal{R}_2^{({\rm m})}$ is less than suggested by a $\sim A_{02}^4$ power-law scaling, which indicates the saturation occurring in the strong-coupling regime. 
Conversely, despite the saturation effect, the bichromatic rate $\mathcal{R}_2^{({\rm bi})}$ has grown more than suggested by a $\sim A_{02}^2$ scaling [see Eq.~\eqref{R2_bi_weak}]. This is because the parameters of the low-frequency field ${\bm{\mathcal{A}}}_1$ have distinctly changed in comparison with Fig.~\ref{figure2}. They now correspond to an enlarged intensity of $\sim 10^{12}$\,W/cm$^2$ and a ponderomotive potential of $U_p\approx 10$--17\,eV. The latter increases the energy threshold for ionization substantially. The absorption of at least $n_0\sim 50$ photons $\omega_1$ from the strong field component, in addition to one high-frequency photon $\omega_2$, is required to reach the continuum.

The saturation effect is also responsible for the circumstance that the relative enhancement factor between the monochromatic two-center and single-center ionization channels, $\mathcal{R}_2^{({\rm m})}/\mathcal{R}_1^{(2)}\sim 10^5$, is less than in Fig.~\ref{figure2}. This is a general result, which was also obtained in \cite{2CPI}: While the absolute magnitude of 2CPI rates is enlarged in the strong-coupling regime, the relative enhancement as compared with the competing single-center ionization channel is reduced. 
Also the ratio of the bichromatic rates has decreased to $\mathcal{R}_2^{({\rm bi})}/\mathcal{R}_1^{(1)}\sim 10^3$. Nevertheless, for the parameters chosen in Fig.~\ref{figure4}, the bichromatic 2CPI rate is still by far the largest.

Regarding the monochromatic single-center PI rate $\mathcal{R}_1^{(0)}$, we note that at least $n_0\sim 85$ low-frequency photons $\omega_1$ are required to overcome the ionization threshold. The Keldysh parameter of $\gamma\sim 0.7$ implies that the coupling with the field ${\bm{\mathcal{A}}}_1$ is situated in the nonperturbative regime of above-threshold ionization.

\begin{figure}[t]
\vspace{-0.25cm}
\begin{center}
\includegraphics[width=0.5\textwidth]{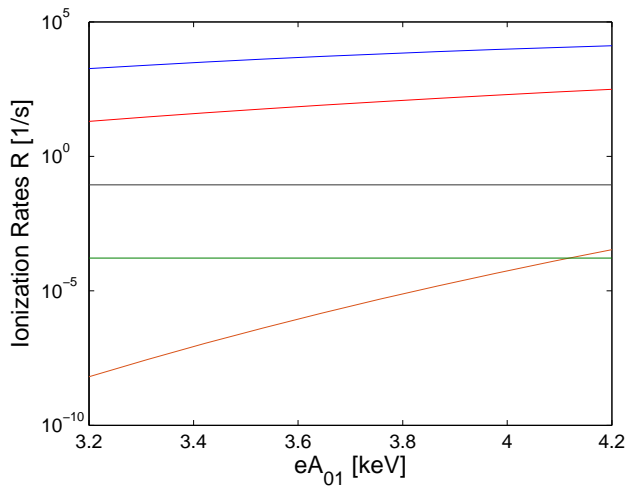}
\end{center}
\vspace{-0.5cm} 
\caption{Photoionization in a He-Ne-like system with $\omega_1=0.85$\,eV, $\omega_2=16.85$\,eV, and $eA_{02}=3.7\times 10^{-2}$\,eV (strong-field case) at interatomic distance $R=5\,\mbox{\AA}$. The various ionization channels are distinguished by the same color coding as in Fig.~\ref{figure2}.}
\label{figure5}
\end{figure}

Figure~\ref{figure5} illustrates the strong-coupling regime for our He-Ne-like model system. For the chosen parameters, the resulting ionization rates quite closely resemble the ones in Fig.~\ref{figure4}, 
exhibiting the same relative order. In particular, the bichromatic 2CPI rate $\mathcal{R}_2^{({\rm bi})}$ is the largest; it dominates over the other rates by at least two orders of magnitude. 
In comparison with Fig.~\ref{figure3}, the rate $\mathcal{R}_2^{({\rm bi})}$ has strongly increased by eight orders of magnitude. The same holds for the monochromatic 2CPI rate $\mathcal{R}_2^{({\rm m})}$, which now lies three orders of magnitude above the single-center rate $\mathcal{R}_1^{(2)}$ in the field $\mathcal{A}_2$. Regarding the single-center rate $\mathcal{R}_1^{(0)}$ in the field $\mathcal{A}_1$, we note that about $n_0\sim 45$ low-frequency photons must be absorbed at least, with the Keldysh parameter $\gamma\sim 1$ indicating the above-threshold regime as before.

\section{Conclusion}
Photoionization of two-center atomic systems in strong laser fields has been considered. The ionization occured through resonant photoexcitation with subsequent radiationless energy transfer to the neighboring atom, combined with additional multiphoton absorption to overcome the ionization threshold. The case of monochromatic fields was treated to establish a direct generalization of earlier studies on 2CPI with single-photon absorption to the multiphoton regime at higher field intensities. Ionization rates large enough to be measured in experiment are difficult to achieve in this scenario, though.

Therefore, the focus was laid on 2CPI in bichromatic fields, consisting of a weak resonant field component and a rather strong low-frequency component which allows for sizeable multiphoton absorption. Various laser-atom interaction regimes were studied. The relative enhancement of strong-field 2CPI over the competing single-center process is particularly high when the coupling to the resonant field is relatively weak ($\Omega_B\ll\Gamma_B$). However, larger absolute 2CPI yields can be achieved in the opposite regime of strong coupling where the resonant field-induced energy shift exceeds the natural line width of the excited state. Also for the low-frequency field component various interaction strengths were analyzed, ranging from the perturbative multiphoton domain to the nonperturbative regime of above-threshold ionization.

Numerical calculations to illustrate the effects were performed on the basis of generic two-center model systems which, despite their relative simplicity, still enable one to capture the essential physics of 2CPI in strong-laser fields. Our general predictions on largely enhanced ionization yields might be tested experimentally by using as real system, for instance, He-Ne dimers in the presence of a weak soft-xuv beam, which is in resonance with a dipole transition in neon, and a moderately strong (near-)optical laser field.

\section*{Acknowledgments}
This study has been performed within the projects MU 3149/4-1 and VO 1278/4-1 
funded by the German Research Foundation (DFG).
We thank R. Baumg\"artel for his contribution at the initial phase of the study.


\end{document}